\documentclass{elsart3}

\usepackage{amssymb}
\usepackage{graphicx} 

\begin{document}

\begin{frontmatter}



\title{Microscopic Derivation of Magnetic Coupling in Ca$_3$Co$_2$O$_6$}


\author{C. Laschinger$^1$, T. Kopp$^1$, V. Eyert$^1$, and
R. Fr\'esard$^2$\corauthref{cor1}} 
\corauth[cor1]{Corresponding author. Tel.: +33 231 45 26 34; fax: +33
231 95 16 00; email address: Raymond.Fresard@ismra.fr}
\address{${}^1$ Institut f\"ur Physik, Universit\"at
Augsburg, 86135 Augsburg, Germany}
\address{${}^2$ Laboratoire Crismat-ENSICAEN, 6, Bld. du Maréchal Juin,
14050 Caen, France}
\begin{abstract}
For cobalt atoms placed along chain structures in Ca$_3$Co$_2$O$_6$, 
we investigate the spin exchange coupling between atoms in high
spin states. Consistent with experimental findings, the coupling 
is weakly ferromagnetic.
\end{abstract}

\end{frontmatter}

Quasi one-dimensional structures have long been acknowledged
to exhibit intriguing transport and magnetic properties. New members
in this family are Ca$_3$Co$_2$O$_6$ \cite{Maignan} and a series of
related iso-structural compounds, which crystallize in the
$R\bar{3}c$ structure \cite{Fjelvag}. In this
structure, two inequivalent cobalt atoms
exist, one in an octahedral environment (labeled Co1), and the other
one in a trigonal-prismatic environment (labeled Co2). Both atoms are
in a 3+ configuration, but in  
different spin states. Cobalt atoms form covalent bonds along chains
while the interchain coupling is far weaker. One
therefore expects that this rather unique situation is going to give
rise to a peculiar magnetic response.

Along the chains the cobalt spins tend to order {\it ferromagnetically}
whereas the interchain coupling is {\it antiferromagnetic}.
However, the antiferromagnetic order is affected by the 
frustration inherent in the crystal structure.
At low  temperature, this system appears to be in 
a partially disordered antiferromagnetic state. On top, the
magnetization versus field curves display several plateaus \cite{Maignan}.

An appealing starting point to the understanding of the magnetic
structure of a particular compound is provided by the
Kanamori-Goodenough-Anderson rules \cite{Goodenough}. Given the
electronic configuration of the ions one can estimate their magnetic
couplings and, at mean-field level, the magnetic ground state. In
Ca$_3$Co$_2$O$_6$, the problem is more involved, in particular since
each second Co$^{3+}$ is 
non-magnetic, as indicated by neutron
scattering data \cite{Fjelvag}. One therefore needs to determine the
next-nearest neighbor magnetic coupling. This is the purpose of the paper.

To that aim one needs i) to determine the electronic configuration of
the ions, and ii) the matrix elements of the hopping operator. Both
are described by a set of parameters obtained from density functional
theory calculations  
using the augmented spherical wave method
(ASW) \cite{Eyert_ASW}. In particular, it turns out that both Co ions are in a
$3d^6$ configuration.
The parameters entering the local part of the standard multi-band
Hubbard Hamiltonian, 
upon integrating out the oxygen degrees of freedom, are the
Hubbard $U$, the Hund's rule coupling 
$J_H$, and the crystal field splitting 10 $Dq$. For Co2, the splitting
of the ``three-fold degenerate'' lowest level turns out to be small
\cite{Eyert}, and is neglected. In contrast to 10 $Dq$ 
both $U$ and $J_H$ are weakly
affected by the difference between the prismatic and octahedral
environments.
The six electrons can
be in three spin configurations: a low spin one (LS) ($S=0$) with
energy $E_{LS} = 
(30 U 
-\sqrt{(3J_H+20Dq)^2 + 24 J_H^2} - 57 J_H - 28Dq )/2$, an
intermediate spin one 
(IS) ($S=1$) with 
energy $E_{IS} = 
(30 U 
-\sqrt{(J_H+20Dq)^2 + 8 J_H^2} - 65 J_H - 8Dq )/2$, or a high spin one
(HS) ($S=2$) with 
energy $E_{HS} = 15 U - 38 J_H - 4 Dq$. As a result the LS
configuration is stable for $J_H<2.68Dq$, and the HS one
otherwise. Stabilizing the IS configuration requires different
physical input, such as an additional splitting of the $e_g$ level or
a lowering of the symmetry \cite{Pouchard}.

We are now in the position to estimate the intra-chain magnetic
couplings. For Co1
we choose the 
coordinate axes to point towards the oxygens, while for Co2
we choose the 
$z$-axis to point along the c-direction of
the $R\bar{3}c$ group. Therefore when determining the matrix elements
of the hopping term one needs to carefully 
distinguish the relevant
contributions. It turns out that the $3z^2-r^2$ orbital on Co2
identically couples to the $x'y'$, $x'z'$, and $y'z'$ orbitals on Co1,
with amplitude $t$, while other matrix elements are negligible. At
this stage one can determine the magnetic ground state performing a
hopping expansion. 
It is obvious that the magnetic structure does not
enter the result to order $t^2$. To order $t^4$ 
the exchange paths between two Co2 sites are sensitive to the
magnetic structure and the respective Hund's rule coupling on Co1.
A typical exchange path is sketched in Fig.~1 for a ferromagnetic
configuration. Any $t_{2g}$ down spin on Co1 can hop onto the
$3z^2-r^2$ orbital on one neighboring Co2, and any remaining $t_{2g}$
down spin on Co1 can hop onto the $3z^2-r^2$ orbital onto the 
second neighboring Co2, (this leaves Co1 on a spin state where $J_H$ is
important), and then both electrons return to Co1. There are 48 such
contributions.
For an antiferromagnetic configuration, 
we find that:
i) there are only 42 such paths, and ii) it leaves Co1
on an intermediate low spin state, which has a higher energy. As a
result the ferromagnetic configuration has a lower energy. 
Using the numerical values
resulting from the ASW calculation \cite{Eyert} and $U=5.5$~eV, 
we finally obtain
\begin{equation}
E_{F} - E_{AF} \simeq -10^{-3} t^4/eV^3  \;.
\end{equation}
Using $t\simeq 0.8$~eV, the energy gain is around 
40~K, in good agreement with the onset temperature of the
ferromagnetic ordering 
along the chains.

\begin{figure}
\centering
\includegraphics[width=0.35\textwidth]{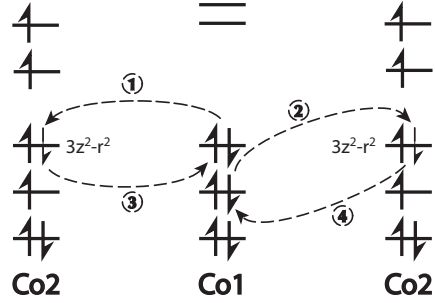}
\caption{Exchange paths, as discussed in the text.}
\label{fig:exchange}
\end{figure}

In summary, we have identified a mechanism leading to a ferromagnetic
coupling between magnetic ions through a non-magnetic one. The driving
forces are i) the Hund's rule coupling on the non-magnetic ion, and ii)
the large number of contributions to the magnetic energy due to the
$R\bar{3}c$ structure. The magnetic coupling is found to be in good
agreement with experiment.

We are indebted to A. Maignan for numerous enlightening discussions.
C. Laschinger is supported by a Marie Curie fellowship of the European
Community program under number HPMT2000-141.  The
project is supported by BMBF~(13N6918/1) and by DFG through SFB~484.

\vspace{- 0.8 cm}

\end{document}